\documentclass[preprint]{elsarticle}
\usepackage{graphicx,subcaption,ragged2e}
\graphicspath{ {images/} }
\usepackage{epstopdf}
\usepackage{hyperref}
\usepackage{amsmath}

\usepackage{color}
\definecolor{green}{rgb}{0,0.5,0}
\usepackage{xcolor}
\usepackage{textcomp}
\def\bel{\begin{equation}\label}
\def\beq{\begin{eqnarray}}
\def\beel{\begin{eqnarray}\label}
\def\eeq{\end{eqnarray}}
\def\be{\begin{equation}}
\def\ee{\end{equation}}
\def\eq{&=&}

\def\bm{\begin{math}}
\def\me{\end{math}}

\def\qq{\qquad}

\def\beel{\begin{eqnarray} \label}

\newcommand \bei {\begin{itemize}}
\newcommand \eei  {\end{itemize}}

\begin{document}

\title{Lane formation in an active particle model with chirality for pedestrian traffic}

\author[1]{Anna S. Bodrova\corref{cor1}}
\ead{bodrova@polly.phys.msu.ru}

\author[2,3]{Fatema Al Najim}
\ead{najimfa@hotmail.com}

\author[4,2]{N. V. Brilliantov}
\ead{nb144@leicester.ac.uk}

 \cortext[cor1]{Corresponding author}

 \affiliation[1]{organization={Moscow Institute of Electronics and Mathematics, HSE University},%addressline={Tallinskaya 34},
                 postcode={123458}, 
                 city={Moscow}, 
                 country={Russia}}

\affiliation[2]{organization={ School of Computing and Mathematical Sciences, University of Leicester},
                 postcode={LE1 7RH}, 
                 city={Leicester}, 
                 country={UK}}
                 
 \affiliation[3]{organization={Department of Mathematics and Statistics, College of Science, King Faisal University}, addressline={P. O. Box 400},
                 postcode={ 31982}, 
                 city={Al-Ahsa}, 
                 country={Saudi Arabia}}

 \affiliation[4]{organization={Skolkovo Institute of Science and Technology},
                 postcode={143026}, 
                 city={Moscow}, 
                 country={Russia}}

%\date{\today}

\begin{abstract} 
We analyze the pattern formation in systems of active particles with chiral forces in the context of pedestrian dynamics. To describe the interparticle interactions, we use the standard social force model and supplement it with a new type of force that reflects chirality. We perform numerical simulations of  two pedestrian flows  moving in opposite directions along a corridor. We observe two dynamic phase transitions that occur for varying number densities of particles and strengths of the chirality force: one  from disordered motion to multi-lane motion and another from multi-lane to two-lane motion. We develop a qualitative theory that describes the demarcation lines for these phase transitions in the phase diagram chirality-density. The results of our analysis agree fairly well with the  simulation data. A comparison with previously reported experimental data has been provided. Our findings may find applications in urban and transportation-planning problems. 
\end{abstract}


\maketitle

\section{Introduction} 

Active matter is a substance composed of active particles that demonstrate motility, that is, "the ability to
exhibit motion and to perform mechanical work at the expense of metabolic energy" \cite{RDAllen}. Active particles can take energy from their
environment and convert it into directed motion \cite{ebreview, activereview, ram2010}. 
Living matter (biological systems) provides  an uncountable amount of examples of  active particles systems \cite{Viswanathan,bioorg,revprx,actbook}. Further examples include Brownian motors (ratchets) \cite{ratchet}, artificial colloidal particles with self-propulsion \cite{spprl, scirep}, etc.  Synthetic colloidal particles can mimic the functions of living matter; this field has developed rapidly over the past few decades \cite{colrev}. Synthetic microswimmers can autonomously achieve directed motion by self-generated chemical \cite{chem} or thermal \cite{term} gradients. One can also mention  examples from  robotics \cite{Brambilla}, biomedicine \cite{WangGao}, and social science, e.g.  \cite{pedbook,HelbRMPh2001, Helbbook,report}.  

A prominent feature of active particle systems is the formation of self-organized coherent
structures, see e.g.
\cite{ram2010,Vicsek,Levine,Bechinger,Marchetti,Carrillo2010,Toner}.
Among these are intriguing swirling patterns emerging in circular motion, when a group of individuals follow one another around an empty core \cite{swirl1,swirl2,swirl3, 
swirl4,BATM2020}. Such vortex arrays and patterns may be formed by different mechanisms, including effective interaction forces between the active particles \cite{ram2010,Levine,Bechinger,Marchetti,Toner, BATM2020}, the intention of animals to escort the center of mass of a swarm \cite{NPB2021}, or intrinsic chirality of the particles, when they can self-propel and self-rotate with a preferable rotation in a certain direction \cite{chir}. 

Here, we focus on the pattern formation in a system of particles with chiral interactions. In this case, the active particles do not possess self-propelling  chirality, as shown in Ref. \cite{chir}, but interact with  forces that have left-right asymmetry.  We will study such systems in the context of pedestrian dynamics, considering pedestrians as active particles with an intrinsic bias in inter-personal interactions.

The mathematical modeling of pedestrian flows, which is an interesting academic problem, has high practical importance  \cite{pedbook,HelbRMPh2001,report}. Indeed, it is directly related to the effective planning of public places, transportation systems, evacuation strategies in emergencies, and the organization of pedestrian fluxes in an epidemic, e.g.  \cite{pedbook,HelbRMPh2001, Helbbook,report,statphys}. There are a variety of approaches to describe pedestrian motion, the most prominent of which are fluid dynamics models, particle models, agent-based models, e.g.   \cite{pedbook, HelbRMPh2001, Helbbook,report, statphys}, and models based on cellular automata \cite{blueadler, capre, Yang2008}. Conceptually, the agent-based approach is similar to the molecular dynamics approach, in which the motion of all molecules is simulated by solving Newtons' laws. Obviously, this is the most microscopic approach that provides the maximum possible information about the system. 

Similar to atomic motion, pedestrian motion can be modelled if one introduces "forces" acting on the pedestrians.  These forces are not real physical forces between human bodies (unless the crowd is dense and people touch each other), but rather reflect the reaction of people on the surrounding neighbors or material objects. In other words, they describe the   {\it intention}  of people to move in a particular direction.  In other words, such forces, which  are called "social forces" \cite{HelbRMPh2001,Helbbook,active} quantify individuals' internal motivation to perform certain movements. They do not describe mechanical interactions but rather reflect the social influence (social norms, public opinion, etc.) on behavioral patterns.

The social force model allows the prediction and description of the motion of pedestrians in different situations: the formation of ordered lanes of pedestrians who intend to walk in the same direction \cite{active}, behavior patterns in the escape panic \cite{hebnature}, motions through the bottleneck \cite{pedbook,HelbRMPh2001,Helbbook}, epidemic dynamics \cite{NBPlosOne}, overtaking behavior of pedestrians \cite{over}, etc. 

Among the basic pattern formation in pedestrian systems is the formation of lanes, comprising pedestrians moving in the same direction. This has been observed for deterministic dynamics without fluctuations \cite{active} and is driven by the propulsion term.  The presence of fluctuations can  prevent lane formation or even destroy existing lanes \cite{hebprl}. A further increase in the intensity of fluctuations may cause the emergence of a crystalline state, the so-called "freezing upon heating" \cite{hebprl}. At high densities, a transition from laminar to turbulent flow can be observed \cite{hebpre2007}

Multi-lane formation has been observed not only in systems of pedestrians, but also in colloidal systems, or pair-ion plasmas in the electric field, where positively  and negatively charged entities (macroions and ions) form lanes of oppositely moving particles, e.g.  \cite{LowenLanes2002,EurJPLane2008,SoftMatt2019,plasma2020}; such self-organization also occurs with increasing particle density. The number of lanes may vary, depending on the number density and other parameters of the system. Band-to-lane
transitions have been observed in discretized Vicsek models \cite{vis1, vis2}, and longitudinal band ("lane") formation has been reported in systems of self-propelled rods and in active nematics \cite{nem1, nem2, nem3, nem4}.

In most previous studies, pedestrian motion was assumed to be symmetric in terms of its reaction to the left and right surroundings. However, people are asymmetric by nature, and handedness is a form of side bias in humans  \cite{Mandal2001,Ida}. Handedness explicitly influences the way in which one perceives the environment and can influence one’s preference to turn towards their favored hand \cite{McBeath,Scharine}. A commonly accepted view of handedness is that approximately 10\%-12\% of humans worldwide are
left-handed \cite{Kushner}. Comparative studies also provide evidence that reinforces 
the view that about 90\% of the population prefer using their right hand to execute unimanual tasks and activities \cite{Mandal2001}. A cross-cultural study surveying artwork spanning over 5,000 years found that 93\% of people appeared to be right-handed, irrespective of
geographical location \cite{coren1977}. Older adults (93.1\%) preferred to use their right hand to
perform tasks \cite{Predtechenskii}. The prevalence of right- and left-handedness varies within approximately 10\% across cultures \cite{Ida,Kushner,Eze}.  

Similar findings have been reported for pedestrian behavior. A study comparing the effects of handedness on side preferences among pedestrians in the USA and  UK showed 
that side preferences are affected by handedness \cite{McBeath,Scharine}. Side preferences exist in daily life in different situations and are closely related to lane formation, as they can affect the type and order of lane formation \cite{Meyers,Takimoto,Jia,Li2015}. 

Investigation of left-right preferences for the counterflow of pedestrians in corridors showed that right-handed pedestrians have a greater preference to walk on the right side of a corridor by evading collisions efficiently \cite{Mu2016}. In Ref. \cite{Yang2008} a cellular automata model was used to study the left-right preference of pedestrians. The authors observed the formation of two lanes with opposite velocity directions and a strong dependence of the total pedestrian flux on the degree of the left-right preference, modelled by the probability bias to move to the right or left cell. In \cite{lftrans,lfphysa} right/left drift rules were introduced as a modification of the driving force. However, the model proposed in \cite{lftrans} does not describe all the aspects of the observed experiments.

%Similar problems have been also investigated in \cite{lfphysa}.

In this study, we consider this natural side preference and introduce a new type of interaction between pedestrians. Namely, we explicitly add the chirality force to the social force and analyze its impact on pattern formation for two opposite pedestrian flows in a long corridor. We demonstrate that with increasing pedestrian density and enhancing chirality force, the system undergoes a couple of dynamic phase transitions: one from disordered motion to motion with several lanes, and another -- from several lanes to two-lane motion. We also demonstrate that our model can qualitatively explain the experimental results of Guo et al \cite{lftrans}.

The remainder of this paper is organized as follows. In the next Section II we specify the model and simulation details. In Sec. III, we discuss the dynamic phase diagram, describing the state of the system. In Sec. IV, we present a qualitative theory of the observed phase transitions. In Sec. V, the results of the motion in a ring-shaped corridor are presented. Finally, in Sec. VI,  we summarize our findings. 

\section{Model} 

The evolution of the system occurs according to the following equations of motion:
\beq \label{drdt}
\frac{d{\bf r}_i}{dt} \eq  {\bf v}_i 
\qquad \qquad \qquad \qq i=1, \ldots N; \\  \label{eqmotion}
M\frac{d\textbf{v}_i}{dt} \eq \textbf{F}^{\rm des}_i+\textbf{F}^{\rm wall}_i+\textbf{F}^{\rm fluc}_i+\textbf{F}^{\rm pp}_i +\textbf{F}^{\rm chir}_{i}, 
\eeq
where $N$ is the total number of pedestrians and  $M$ is the pedestrian mass. %In what follows we will use the units with $m=1$. 
In Eq. \eqref{eqmotion} $\textbf{F}^{\rm des}$ is the destination (propulsion) force, describing the tendency of a pedestrian to move with a certain  (desired)  velocity ${\bf v}^{\rm des}$:
\bel{Fdes}
\frac{\textbf{F}^{\rm des}_i}{M} = -\tau^{-1}\left( {\bf v}_i - {\bf v}^{\rm (des)}_i \right). \ee
%
%\begin{eqnarray}
%F^{\rm des}_x=\frac{-v_x+v_{\rm des}}{\tau}\\
%F^{\rm des}_y=-\frac{v_y}{\tau}
%\end{eqnarray}
Here  ${\bf v}_i$ is the velocity of $i$-th pedestrian, ${\bf v}^{\rm des}$ is the desired velocity and $\tau$ is the respective relaxation time; here we use $\tau=0.5\,[s]$ and $|{\bf v}^{\rm (des)}|=1.34\,[m/s]$.  

$\textbf{F}^{\rm wall}$  describes the interaction between a pedestrian and walls, columns, and other objects. We used the following expression for this force:
\begin{eqnarray}
\label{wall}
F^{\rm wall}_{x,i}&=&0,\\
\frac{F^{\rm wall}_{y,i}}{M}&=&\frac{U_0}{\Delta L}\left[\exp\left(-\frac{y_i}{\Delta L}\right)-\exp\left(\frac{y_i-L_y}{\Delta L}\right)\right], 
\end{eqnarray}
with $U_0=10\,[m^2/s^2]$ and $\Delta L=0.2\,[m]$; that is, we consider the motion in a corridor, with the walls parallel to the direction of motion ($x$-direction), located at $y=0$ and $y=L_y$. 

$\textbf{F}^{\rm fluc}$ refers to the fluctuation force. Here, we use for $\textbf{F}^{\rm fluc}$  a  white Gaussian noise with zero mean and dispersion $\sigma^2/M^2=0.01\, [m^2/s^4]$. Such noise intensity corresponds to the case when the fluctuation force comprises a few percents of the destination force.  

Let $\bf{r}_{ik}=\left(\textbf{r}_i-\textbf{r}_k\right)$ be the vector connecting the centers of $i$-th and $k$-th pedestrians, and $\textbf{e}=\bf{r}_{ik}/d$ be the corresponding unit vector, with $d=\left|\bf{r}_i-\bf{r}_k\right| $ being the distance between the pedestrian centers.
%$$d=\sqrt{\left(x_k-x_i\right)^2+\left(y_k-%y_i\right)^2}$$  
%$$d=\left|\bf{r}_i-\bf{r}_k\right|$$  
%Then $\textbf{F}^{\rm pp}$ describes interactions between pedestrians. It depends on the inter-pedestrian distance and their velocities. 
We consider the circular specification of the social force, which describes the interactions between pedestrians \cite{HelbRMPh2001,Helbbook}: 
\begin{eqnarray}
\label{Fpp}
&&{\bf F}^{\rm pp}_i =  \sum_{k \neq i} {\bf F}^{\rm pp}_{ik}, \\ \nonumber
&&\frac{{\bf F}^{\rm pp}_{ik}}{M} = \frac12 A^{-\frac{\epsilon}{B}}\textbf{e}_{ik}\left(1-\textbf{e}_{ik}\cdot\textbf{c}_i\right)
\end{eqnarray}
%\label{Fpp2}
%\frac{F^{\rm pp}_{x,ik}}{m}\eq Ae^{-\frac{\epsilon}{B}}\frac12\frac{x_i-x_k}{d}\left(1-c_{xi}\frac{x_i-x_k}{d}-c_{yi}\frac{y_i-y_k}{d}\right)\,\\\nonumber
%\label{Fpp3}
%\frac{F^{\rm pp}_{y,ik}}{m}\eq Ae^{-\frac{\epsilon}{B}}\frac12\frac{y_i-y_k}{d}\left(1-c_{xi}\frac{x_i-x_k}{d}-c_{yi}\frac{y_i-y_k}{d}\right)\, 

Here the parameter values are: $B=0.3\,[m]$, $A=2.1\,[m/s^2]$.  $\textbf{c}_i=\textbf{v}_i/v_i$ is the dimensionless velocity of particle $i$, specifying its direction, and  $\epsilon=d-2R$, where $R=0.2\,[m]$ is the characteristic "size" of pedestrians. 

%${\bf r}_i=(x_i,y_i)$ and ${\bf r}_k=(x_k,y_k)$ corresponds to the positions of $i$-th and $k$-th pedestrian, $d$ is the distance between their centers,
%$$d=\sqrt{\left(x_k-x_i\right)^2+\left(y_k-y_i\right)^2}$$  
%and  $\epsilon=d-2R$, where $R=0.2\,m$ is the characteristic "size" of pedestrians. 
%$$c_{xi}=v_{xi}/\sqrt{v_{xi}^2+v_{yi}^2}, \qquad   c_{xi}=v_{yi}/\sqrt{v_{xi}^2+v_{yi}^2}$$ 
%are the dimensionless components of the unit vector ${\bf c}$, specifying the direction of the velocity of the  $i$th pedestrian.  
In what follows we will analyze the kinetic behavior of a system of pedestrians, as a system of "particles" with specific interaction forces. That is, although the term "particles" below mainly refers to pedestrians, the results may be applicable to other systems of active particles with the same forces.

\begin{figure*}\centerline{\includegraphics[width=1.8\textwidth]{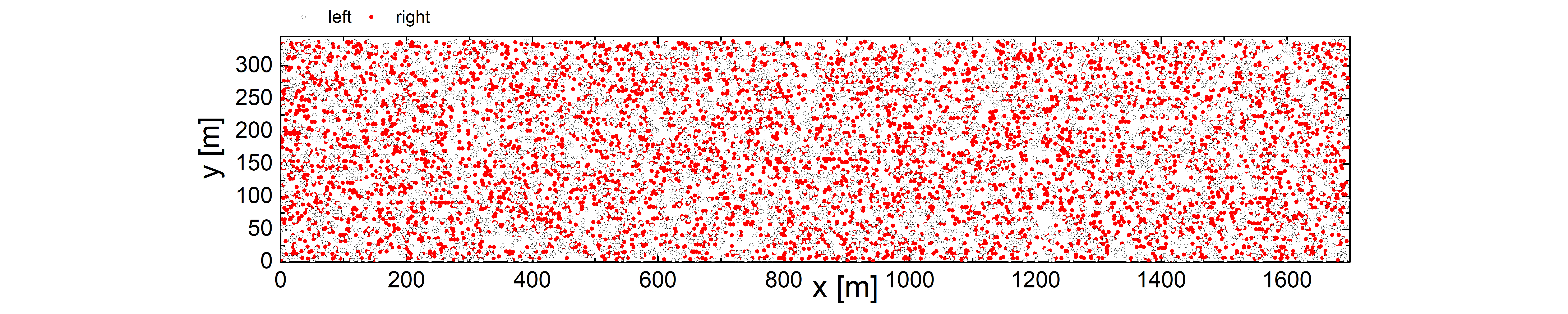}}\caption{The disordered motion of active particles at low number density $\rho=0.02\,[m^{-2}]$ and chirality $\chi=0.001\,[m/s^2]$. Red circles show particles moving to the right and white circles - to the left. For the better visibility the size of particles in this and all other figures is significantly exaggerated. }\label{Gchaotic}\end{figure*}

We assume that a chirality force emerges if the particles (pedestrians)  move in opposite directions and approach each other. Moreover, we assumed that the direction of the force is perpendicular to the particle velocity and that it acts to the right with respect to the direction of motion. Thus, we used the following force: 
\begin{equation}
\label{chir}
\frac{\textbf{F}^{\rm chir}_{ik}}{M}=\chi \Theta\left(D-d\right) \Theta( -\textbf{v}_i \cdot  \textbf{v}_k) \Theta( -\textbf{r}_{ik} \cdot \textbf{v}_{ik}) \textbf{N}_i, 
\end{equation}
%where $\textbf{r}_{ik}=\textbf{r}_i-\textbf{r}_k$ and $\textbf{v}_{ik}=\textbf{v}_i-\textbf{v}_k$ are respectively the inter-particle distance and velocity for the pair $(i,k)$,
Here, $\textbf{v}_{ik}=\textbf{v}_i-\textbf{v}_k$ is the relative velocity of particles $i$ and $k$, $\Theta\left(x \right)$ is the unit Heaviside step function, and $ \textbf{N}_i$ is the unit vector that determines the direction of the chirality force: 
$$
\textbf{N}_i = \left[\textbf{v}_{i} \times \textbf{e}_z  \right] /\left| \textbf{v}_{i} \right|, \qquad {\rm if} \qquad {\bf v}_i \neq 0,  
$$
where $ \textbf{e}_z$ is the unit normal to the plane of motion (ground), pointing upward. 
The first step-function on the right-hand side of Eq.~(\eqref{chir}) shows that a chirality force emerges if the distance between the particle centers $d$ is smaller than $D$. The force is equal to $\chi$ and does not depend on the distance. In our simulations, we used $D=4\,[m]$. The second step function guarantees that the particles move in opposite directions and the third step function guarantees that the particles approach each other. The chirality force in Eq.~(\eqref{chir}),  has the simplest form with the characteristic force quantified by $\chi$ and characteristic length quantified by $D$.  Such a simple dependence may be also motivated by its nature -- it is based on the visual contact, in contrast to the exponential repulsive force, associated with the "physical" body contact. The full chirality force is calculated as the sum of the interactions with all possible particles: $\textbf{F}^{\rm chir}_{i}=\sum\textbf{F}^{\rm chir}_{ik}$.

The force  $F^{\rm chir}_{ki}$ has the same form as \eqref{chir}, but with the vector $\textbf{N}_k = \left[\textbf{v}_{k} \times \textbf{e}_z  \right] /\left| \textbf{v}_{k} \right|$ instead of $\textbf{N}_i$. The positive chirality, $\chi>0$,  corresponds to the chirality force acting to the right, with respect to the direction of motion (associated with  right-handed traffic). The negative chirality,  $\chi<0$, corresponds to this force, acting to the left (associated with left-handed traffic). 

The system was initialized by randomly inserting active particles into $L_x\times L_y$ box without overlaps with the alternating sign of the destination velocity, $v_{\rm des,\,x}= \pm 1.34\,[m/s]$,  $v_{\rm des,\,y}= 0\,[m/s]$. The lateral component of the desired velocity is positive for half of the particles and negative for the other half. The total  number of particles in the simulations (unless mentioned)  is $N=11520$. We analyzed the system with aspect ratio $L_x/L_y=5$ until specified otherwise and applied periodic boundary conditions in the $x$-direction. Then, the system evolved in accordance with the equations of motion \eqref{drdt}-\eqref{eqmotion} until a steady state was established. 

\begin{figure*}\centerline{\includegraphics[width=1.8\textwidth]{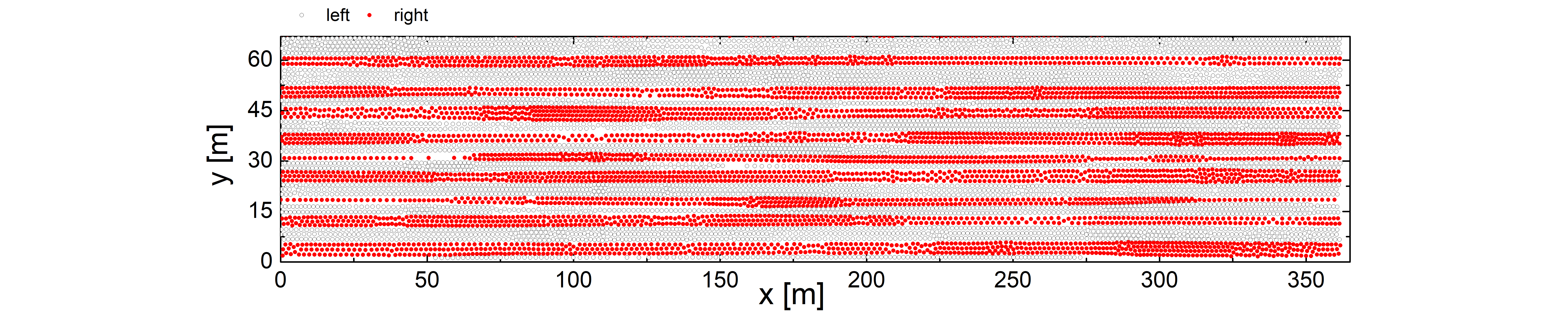}}\caption{Multiple lane formation for the system with very small chirality,  $\chi =0.001\,[m/s^2]$ and relative large number density $\rho= 0.44\,[m^{-2}]$.} %For better visibility the size of the particles is exaggerated.
\label{Gstripes}\end{figure*}

\begin{figure*}\centerline{\includegraphics[width=1.8\textwidth]{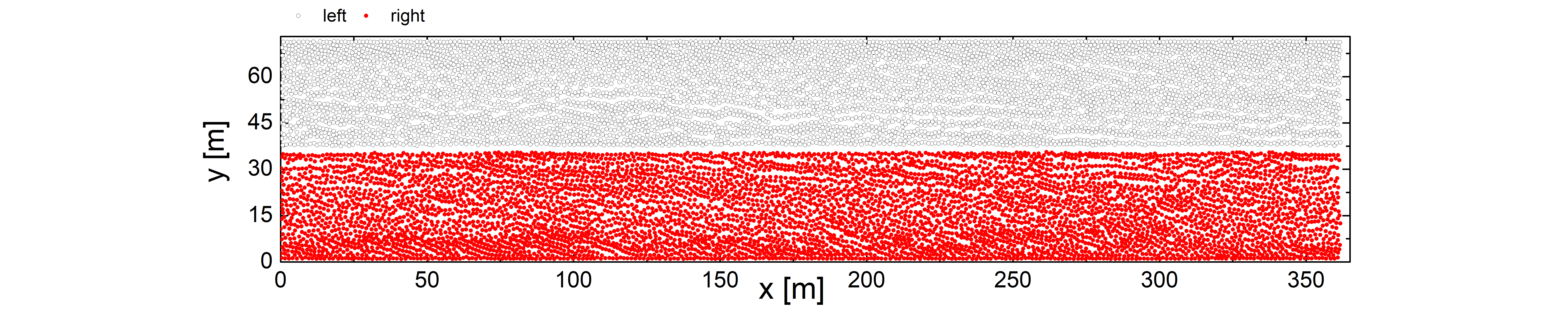}}\caption{Two-lane formation  in the system with $\rho =0.44\,[m^{-2}]$ and  $\chi=0.15\,[m/s^2]$. Two opposite fluxes of active particles spontaneously emerge. } \label{Gtwo}\end{figure*}

\section{Phase diagram} 

Figs. \ref{Gchaotic} -- \ref{Gtwo} demonstrate typical patterns observed in the system. For a small number density of particles, $\rho=N/(L_x \cdot L_y)$, when the system is very dilute, pedestrian motion appears to be disordered, as shown in Fig. \ref{Gchaotic}. For larger values of number density, the system self-organizes into several lanes,  comprised of particles moving in the same direction (see Fig.~\ref{Gstripes}). For still larger values of $\rho$ and non-zero chirality $\chi$, the multi-lane patterns transform into two-lane patterns, that is, into two opposite fluxes, as shown in Fig. \ref{Gtwo}. The simulations of  smaller systems with the same aspect ratio show the same patterns for the same set of parameters.

\begin{figure}
\centerline{\includegraphics[width=0.8\textwidth]{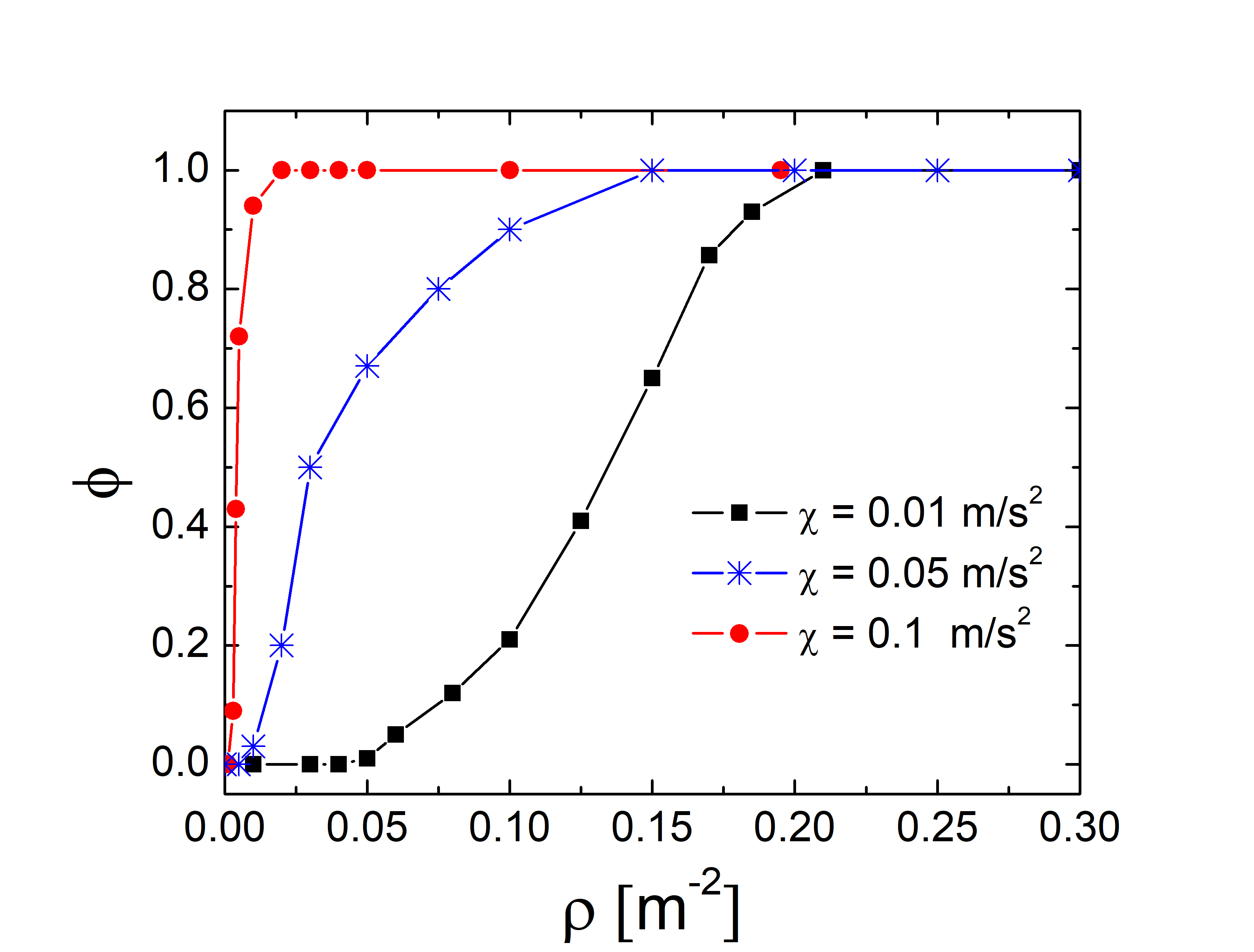}}
\caption{The dependence of the order parameter $\phi$ on the density $\rho$ for different chiralities,  $\chi = 0.01, 0.05, 0.1 \,[m/s^2]$ and aspect ratio $1:5$. With the increasing $\rho$ system undergoes a transition form the disordered state ($\phi=0$) to the two-lane state ($\phi=1$), through the multi-lane state ($0 < \phi < 1$).  With the decreasing  chirality the transition becomes smoother.}\label{Gphi}\end{figure}

To define the state of the system, we introduce an order parameter. We do this in a  similar way as in the corresponding study of binary colloidal mixtures \cite{LowenLanes2002}. Let us assign parameter $\phi_i$ to every particle $i$ moving to the left. We introduce the effective radius of the free space around each particle as
\begin{equation}
r_{\rm min} = \frac{1}{\sqrt{2\rho}}.
\end{equation}
If the distance (in the direction perpendicular to the direction of motion) of particle $i$ and at least one particle $k$ moving to the right is smaller than $r_{\rm min}$, that is,  $|y_i-y_k|<r_{\rm min}$, the parameter $\phi_i$ is set to zero. Otherwise, it was equal to unity. Then, we calculate the average value of $\phi_i$ over all $N_{\rm left}$ particles moving to the left:
\begin{equation}
\phi_{\rm left} = \frac{1}{N_{\rm left}}\sum_{i = 1}^{N_{\rm left}}\phi_i .
\end{equation}
Analogously, we compute $\phi_{\rm right}$ for the particle, moving to the right. The order parameter, which defines the state of the system, is calculated as the average value:
\begin{equation}
\phi = \frac12\left(\phi_{\rm left}+\phi_{\rm right}\right) .
\end{equation}
For completely disordered systems, the order parameter is equal to zero, that is, $\phi = 0$. In the case of the two lines, it becomes equal to unity, $\phi = 1$. For several lanes, intermediate non-zero values can be obtained. The dependence of the order parameter on number density is shown in Fig.~\ref{Gphi}. The transition becomes smoother with decreasing chirality $\chi$.

\begin{figure}\centerline{\includegraphics[width=0.8\textwidth]{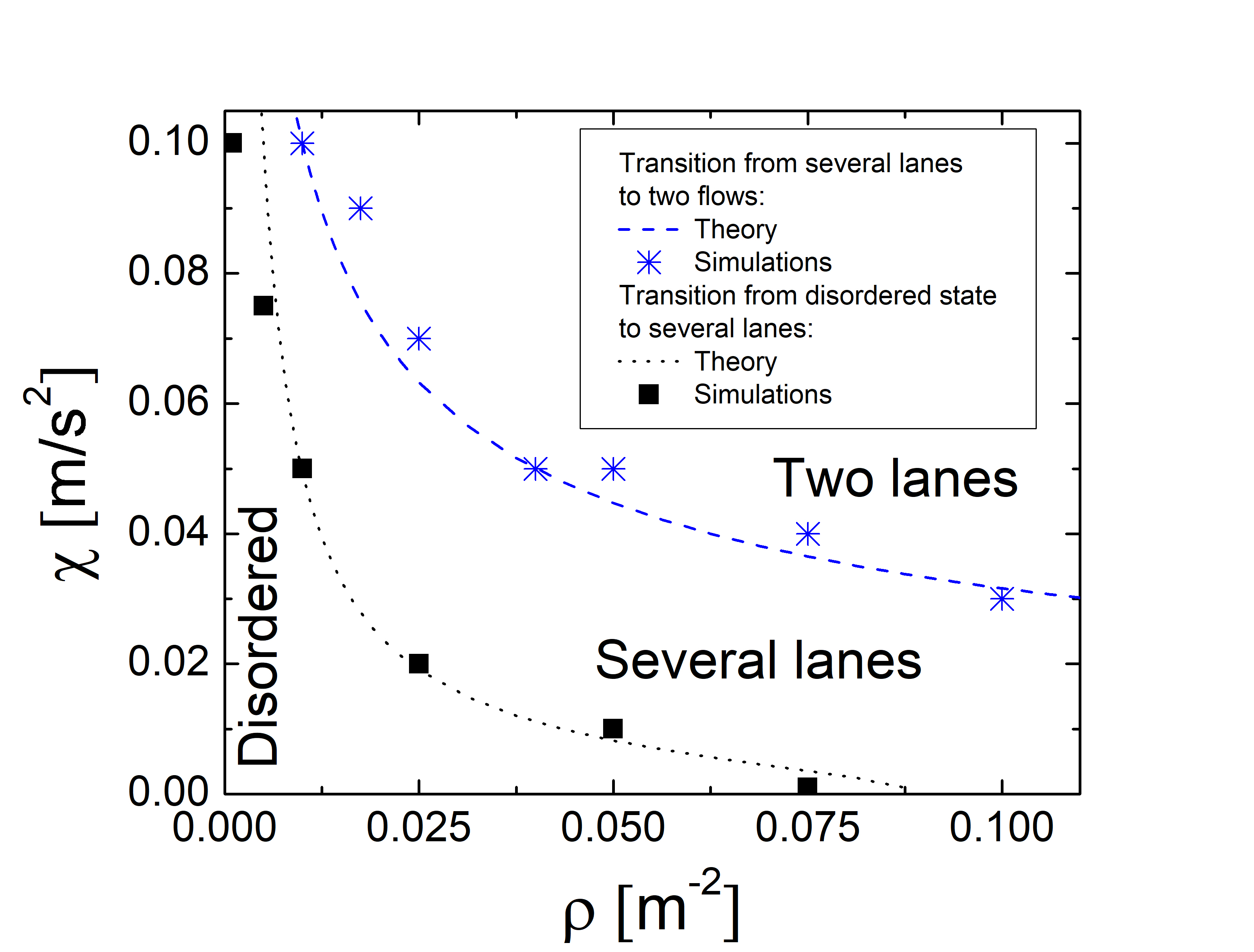}}\caption{The kinetic phase diagram of the system. For very low number density, the  pedestrian motion  is disordered. At larger number density,  a transition from the disordered to multi-lane regime is observed. For still larger density,  the multi-lane pattern transforms  into a two-lane one.  
 Symbols correspond to the  simulation results, lines -- to the prediction of the qualitative theory for with the transition lines $\chi_* (\rho)$ and $\chi_{**}(\rho)$, given, respectively,  by Eqs.~\eqref{trline} and \eqref{chiro2tomu}. The 
 fitting parameters are $C = 4$ for  $\chi_* (\rho)$ and  $Q = 0.01\,[s^{-2}]$ for $\chi_{**}(\rho)$.
 %, Eq. \eqref{chirho}.
 }\label{Gpd}\end{figure}

Let us consider the phase diagram of the system (Fig.~\ref{Gpd}), where the  possible states of the system are given in terms of number density $\rho$ and chirality $\chi$.  The dots correspond to the simulation results, whereas the lines correspond to the results of qualitative theory.  At low number densities, the system resides in a disordered state. With increasing  number density, multi-lane  formation occurs. At a sufficiently large chirality, the increase in the number density leads to a second transition from a multi-lane to a two-lane pattern.

\begin{figure}
\centerline{\includegraphics[width=0.8\textwidth]{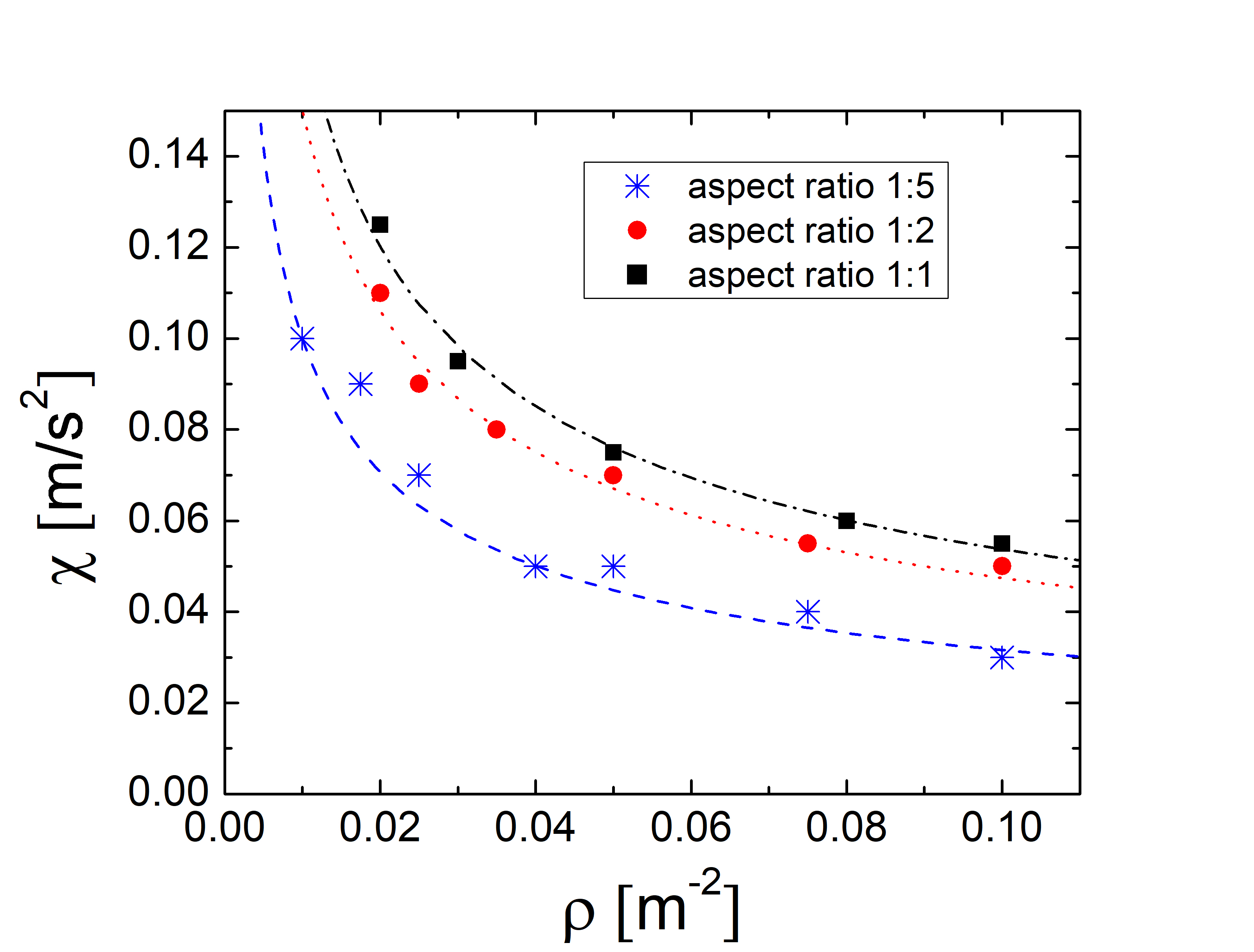}}
\caption{The transition between the two-lane  and multi-lane patterns, for systems with different aspect ratio. Dots correspond to the simulation results, lines - to the qualitative theory, Eq.~\eqref{chiro2tomu}, for $\chi_{**}(\rho)$. The fitting parameters are $Q = 0.01\,[s^{-2}]$ for the aspect ratio 1:5, $Q = 0.015\,[s^{-2}]$ for 1:2 and $Q = 0.017\,[s^{-2}]$ for 1:1.} 
\label{Gaspectratio}
\end{figure}

We also studied the influence of the transition between two lanes and the multi-lane state in systems with
other aspect ratios, such as 1:1 and 1:3. The results are shown in Fig. \ref{Gaspectratio}. As can be seen from the figure, the qualitative behavior of the transition line, demarcating the multi-lane and two-lane patters, persists for other aspect ratios.

\begin{figure*}\centerline{\includegraphics[width=1.8\textwidth]{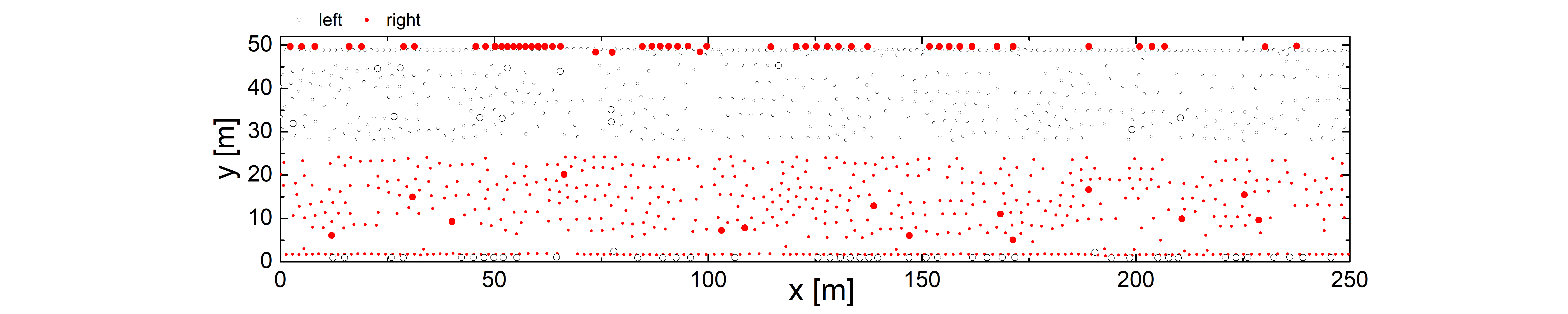}}\caption{Pattern formation in the system, comprised of $90\%$ of right-handed pedestrians with  $\chi=0.1\,[m/s^2]$, and  $10\%$ of left-handed pedestrians (shown with bigger symbols) with  $\chi=-0.1\,[m/s^2]$. The two-lane regime of the system, with the same density and $100\%$ of right-handed pedestrians with the same  $\chi=0.1\,[m/s^2]$, transforms in this case into two-lane regime, with the additional narrow lines. These lanes of oppositely moving particles are located near the walls and contain only pedestrians with negative chirality. The number of particles is  $N=1280$.} \label{Ghir7680}\end{figure*}

 Finally,  we consider the case where $10\%$ of left-handed and $90\%$ of right-handed pedestrians are  present in the system for parameters  $\rho$ and $|\chi|$, corresponding to the two-line state for the case of $100\%$ right-handed pedestrians (here, we assume that for right-handed pedestrians, $\chi>0$, and for left-handed pedestrians, $\chi<0$). The presence of left-handed pedestrians does not destroy the two-lane regime, although, in the vicinity of the walls, the  additional narrow lines appear: The narrow lane  of the "left-handed" pedestrians, near the wide lane of the "right-handed" pedestrians, and vise versa -- a narrow lane of the "right-handed" pedestrians, near the wide lane of the "left-handed" ones, see Fig.~\ref{Ghir7680}. %The presence of negative chirality also leads to spontaneous symmetry breaking: some of particles with positive desired velocity start moving to the left and visa versa. More detailed study of this phenomena will be the subject of future investigations.

\section{Qualitative theory} 

Below, we provide a qualitative analysis of observed kinetic phase transitions. We make mean-field estimates of the forces acting on the system. 

The average force acting between particles can be estimated by its marginal value in the disordered state, that is, by its value in the disordered state at the  transition density. To estimate $\left< F^{\rm pp} \right>$ we used Eq. \eqref{Fpp}, taking particle $i$ at $(x_i,y_i)=(0,0)$ and assuming that it moves with velocity ${\bf v}_i=(v_{xi},0)$. Particle $k$ (representing all the other particles) is uniformly distributed with density $\rho$. This yields the estimate for the $x$-component of the average force: 

\beel{Fppav} 
\left< F_x^{\rm pp} \right>\!\!&\!\approx\!& \!\! -\rho \!\int_0^{2\pi}\!\!d\varphi \int_0^{\infty } \!\!r dr  \tfrac12 A e^{-(r-2R)/B} \cos \varphi (1+ \cos \varphi) \nonumber \\ 
&=& - \tfrac12  \pi   A B^2  e^{2R/B} \rho  = -q \, \rho , 
\eeq
where $q=1.126\,[m^3s^{-2}]$ and we use $(x_i-x_k)/d= (x_i-x_k)/r= -\cos \varphi$ and the above values for parameters $A$, $B$ and $R$. Obviously, the average $y$-component of the force, $F_y^{\rm pp}$, is zero.  

The average chirality force $\left< F^{\rm chir} \right> $, acts in $y$-direction, may be estimated as follows:
\bel{chirav}
\left< F_y^{\rm chir} \right> \approx C \pi D^2  \, \rho \, \chi .
\ee 
Here, $\pi D^2 $ is the area where the chirality force may act, and $\pi D^2 \rho$ is the number of pedestrians in this area. The coefficient $C$ accounts for the geometry of the space and for the fraction of pedestrians (that  approach and move in the opposite direction) that interact with the force $\chi$. The mean field estimate of the total force, which is the sum of the two perpendicular forces, is expressed as follows: 
\bel{fsum}
\left< F^{\rm tot}\right> = \sqrt{\left< F_x^{\rm pp} \right>^2+ \left< F_y^{\rm chir} \right>^2} = \rho \sqrt{q^2 + C^2\pi^2 D^4 \chi^2}.
\ee 
At the same time the mean square value of the stochastic force reads, 
\bel{msqF}\sqrt{\left< (F^{\rm fluc})^2 \right>}/M = \sqrt{\sigma^2}/M =0.1\,[m/s^2]
\ee
We assume that, on average, the stochastic force, which counteracts lane formation,  would destroy the lines, transforming ordered patterns into disordered ones, when the two forces in Eqs.~(\eqref{fsum}) and (\eqref{msqF}) are equal. This yields the following relationship between the number density and chirality on the transition line from a disordered state to an ordered state: 

\bel{trline}
\chi_*(\rho)=(C\pi D^2)^{-1} \sqrt{(\sigma/\rho)^2 -q^2}, 
\ee
This also shows that there exists a threshold intensity of noise; if it exceeds the threshold, the ordered patterns may be destroyed. 

To obtain the criterion for the transition from multi-lane to two-lane patterns, we argue as follows. Let the average distance between successive particles moving in the same direction in the same lane be $l_1$ and the average distance between two neighboring lanes, where particles move in the same direction, be $l_2$. Because the average free area per particle is $1/\rho$, the average distance between the particles scales as $\sim 1/\sqrt{\rho}$. Hence, both $l_1$ and $l_2$ also scale as $\sim 1/\sqrt{\rho}$ and can be written as
\bel{l1l2}
l_1= \frac{C_1}{\sqrt{\rho}}\,, \qquad \qquad l_2= \frac{C_2}{\sqrt{\rho}}\,,
\ee
where the dimensionless parameters $C_1$ and $C_2$ depend on the geometry of the space, that is, the aspect ratio $(L_x/L_y)$ of the corridor. If a particle moves in $x$-direction inside a lane of same-direction particles,  it does not collide with other particles. However, it experiences the action of the chirality force, $\left< F_y^{\rm chir} \right>$ (Eq.~\eqref{chirav}). This drives the particle in $y$-direction perpendicular to its motion. In other words, it drives particles  out of the lane. When a particle moves out of its lane, it may collide with other particles from the  neighboring lane, moving in the opposite direction. Such collisions return the particles to their initial lanes. If the motion in the $y$-direction is very fast, owing to the large chirality force, $\left< F_y^{\rm chir} \right>$, it may happen that the particle will be able to move, without a collision, to the next lane of particles, moving in the same direction, separated by the distance $l_2$. In this case, the action of the chirality force results in merging of  two lanes of same-direction particles, which eventually will result in the formation of a two-lane pattern.  

The average free time between collisions $\tau$, may be estimated as $\tau=l_1/v^{\rm des}$. With acceleration $a=\left< F_y^{\rm chir} \right>/M$, acting during the average time $\tau$, the particle shifts on average in the $y$-direction by $\tfrac12 a \tau^2$. When this shift is equal to $l_2$, the system undergoes, as explained above, a transformation from a multi-lane to a two-lane structure. Hence, the transformation condition $\tfrac12 a \tau^2 =l_2$, 
 reads, 

\bel{twomulti}
\frac12 \frac{\left< F_y^{\rm chir} \right> l_1^2}{M(v^{\rm des})^2} = l_2,
\ee
or, using Eq. \eqref{chirav} for the chirality force, and Eq. \eqref{l1l2} for $l_1$ and $l_2$, we obtain, 
\bel{chiro2tomu}\chi_{**}(\rho)=\frac{Q}{\sqrt{\rho}} ,\ee
%\bel{chiro2tomu}\chi_{**}(\rho)=\frac{Q}{\sqrt{\rho}} \sim \frac{1}{\sqrt{\rho}},\ee
where $Q=2M(v^{\rm des})^2 C_2/(\pi C\,C_1^2 D^2)$. 

In Figs. \ref{Gpd} and \ref{Gaspectratio}, we compare the predictions of our qualitative theory for the transition lines $\chi_* (\rho)$ and $\chi_{**}(\rho)$, given by Eqs. \eqref{trline} and \eqref{chiro2tomu} the simulation results. A fair agreement is observed between the theory and simulations. 

%\begin{figure*}\centerline{\includegraphics[width=0.4\textwidth]{Gcircle0.jpg}\includegraphics[width=0.4\textwidth]{Gcircle5.jpg} \includegraphics[width=0.4\textwidth]{Gcircle15.jpg}\includegraphics[width=0.4\textwidth]{Gcircle40.jpg}}\caption{Motion of $N=100$ pedestrians in a ring-shaped corridor with inner radius $R_1=2\,m$ and external radius $R_2=5\,m$ at different time moments $t = 0, 5, 15, 40 \, [s]$. The pedestrians in the outer line depicted with white circles are moving counterclock-wise, the pedestrians in the inner line (red circles) are moving clockwise. The results of Guo et al. \cite{lftrans} are qualitatively reproduced. } \label{Gring}\end{figure*}

\begin{figure}
\captionsetup[subfigure]{justification=Centering}
\begin{subfigure}[t]{0.55\textwidth}
    \includegraphics[width=\textwidth]{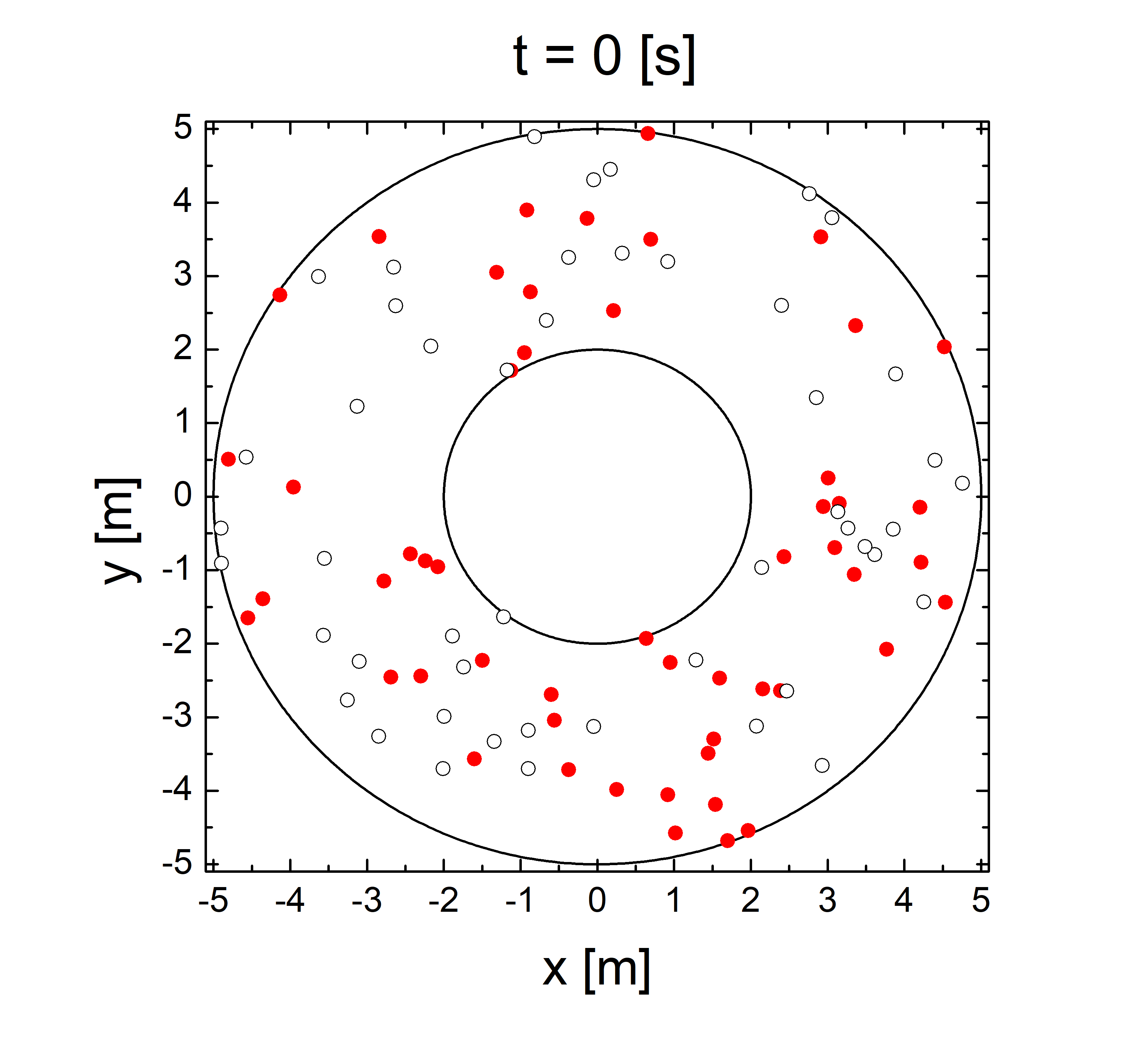}
\end{subfigure}\hspace{\fill} % maximize horizontal separation
\begin{subfigure}[t]{0.55\textwidth}
    \includegraphics[width=\linewidth]{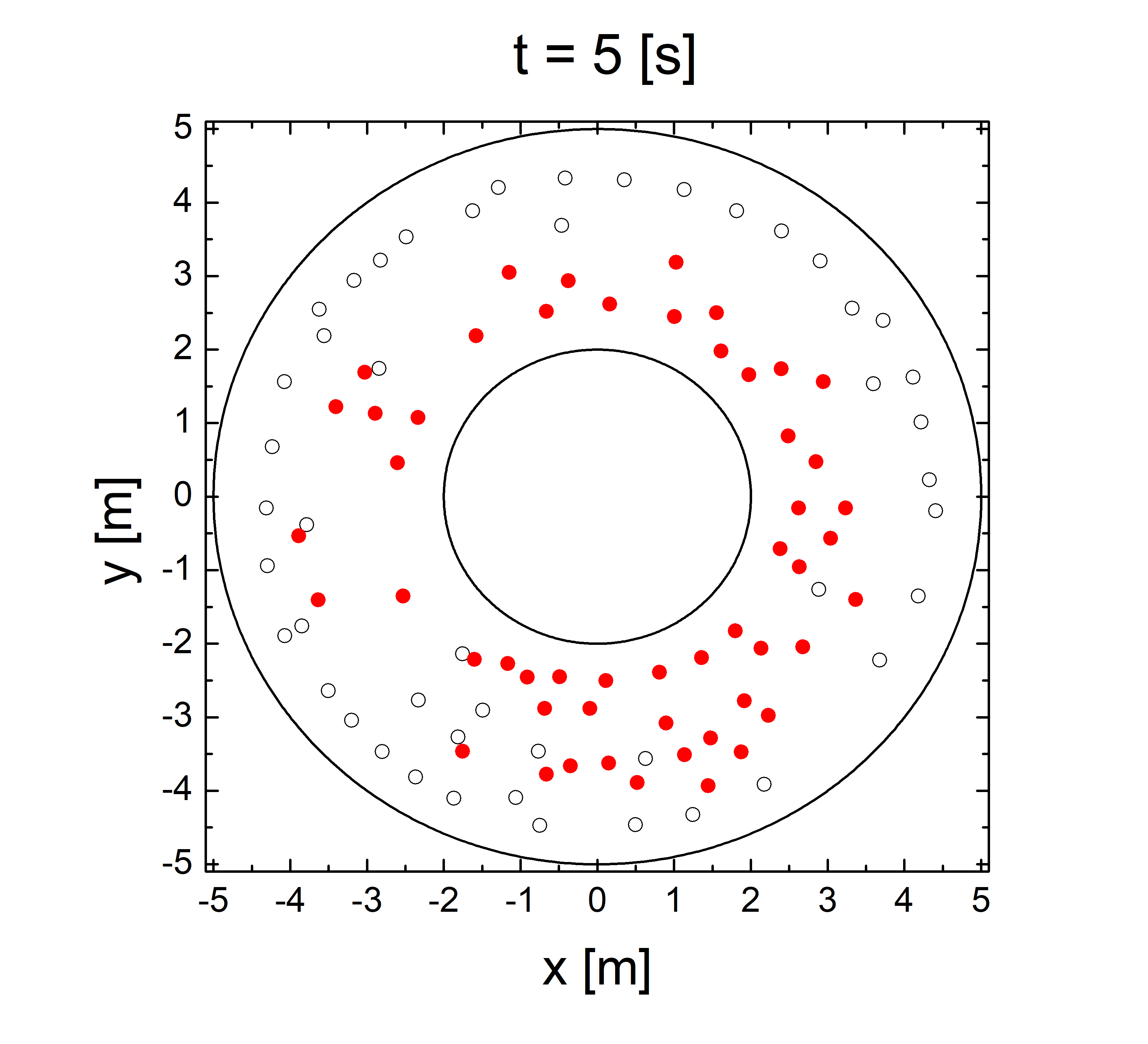}
\end{subfigure}
\bigskip % more vertical separation
\begin{subfigure}[t]{0.55\textwidth}
    \includegraphics[width=\linewidth]{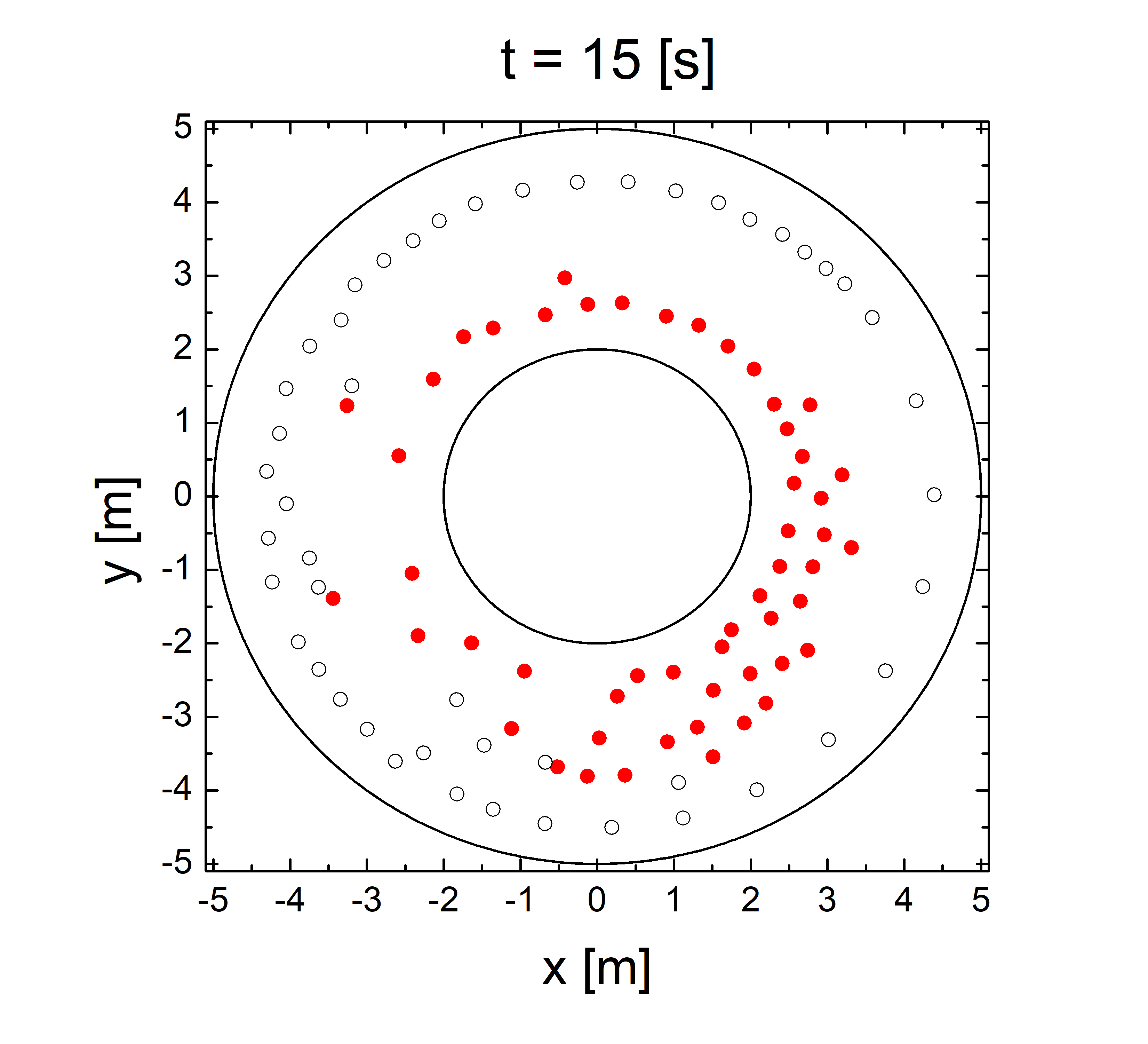}
\end{subfigure}\hspace{\fill} % maximize horizontal separation
\begin{subfigure}[t]{0.55\textwidth}
    \includegraphics[width=\linewidth]{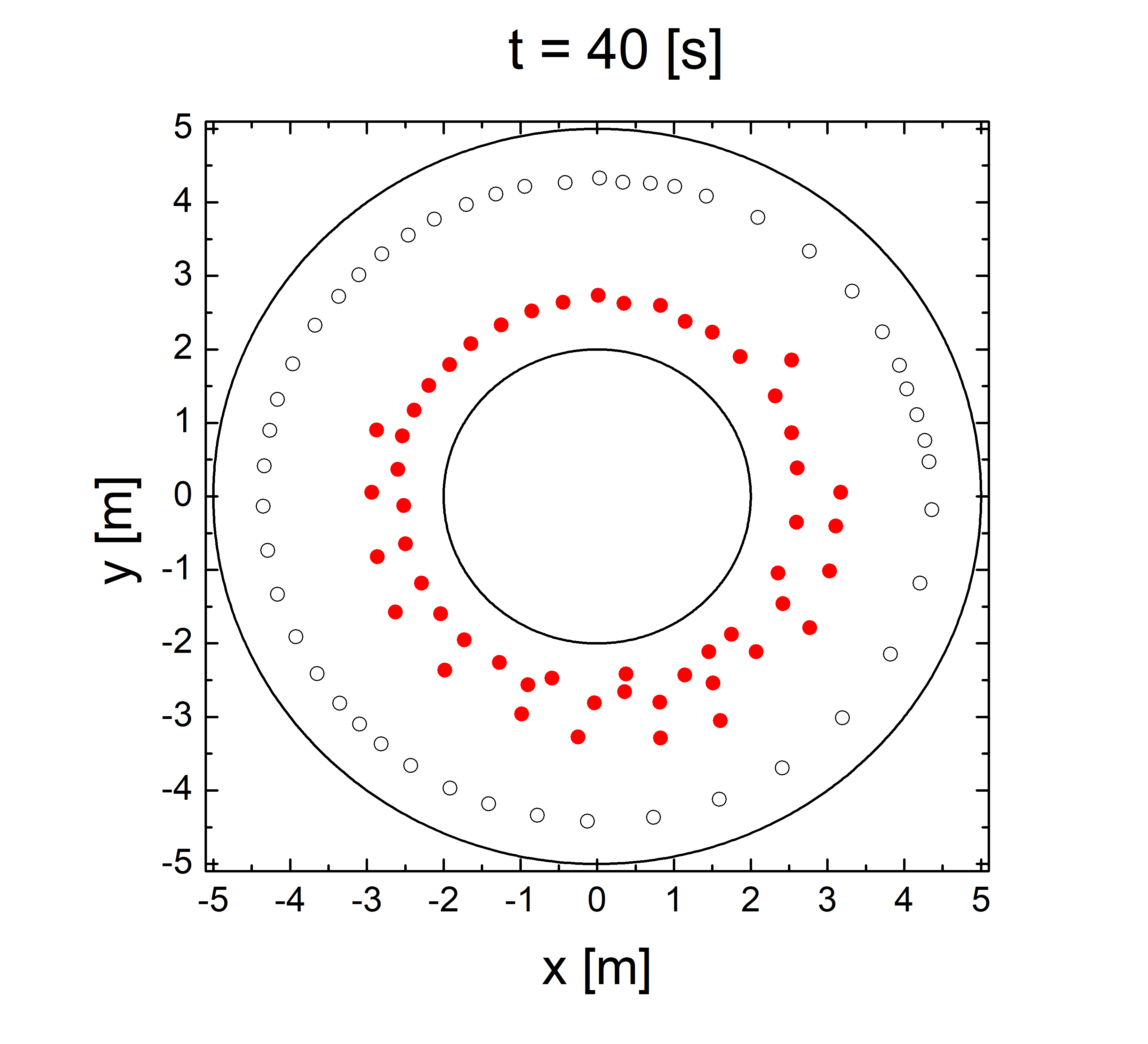}
\end{subfigure}
\caption{Motion of $N=100$ pedestrians in a ring-shaped corridor with inner radius $R_1=2\,m$ and external radius $R_2=5\,m$ at different time moments $t = 0, 5, 15, 40 \, [s]$. The pedestrians in the outer line depicted with white circles are moving counterclock-wise, the pedestrians in the inner line (red circles) are moving clockwise. The results of Guo et al. \cite{lftrans} are qualitatively reproduced. }\label{Gring}
 \end{figure}

\section{Comparison with experiments in the ring-shaped corridor}

We now use our model to describe the experiments conducted by Guo et al.  \cite{lftrans}. They investigated 40-100 pedestrians moving in a ring-shaped corridor with an inner radius $R_1=2\,[m]$ and external radius $R_2=5\,[m]$. In the corridor the spontaneous formation of two lanes has been observed. Due to the right walking preference, the pedestrians in the outer lane always performed the counterclock-wise motion, and pedestrians in the inner lane moved clockwise. This is also qualitatively confirmed in our simulations with $N=100$ pedestrians, see Fig.~\ref{Gring}.  Without the chirality force the probability of the establishment of the  counterclock-wise flow in the inner lane and clockwise flow in the outer lane is the same as the probability of the establishment of the opposite motion, depicted in Fig.~\ref{Gring}. 

Initially, all the pedestrians were randomly placed in the corridor. The transition time was estimated in the experiments as $t = 15\, [s]$ \cite{lftrans}, while in our simulations, the flows were almost separated at this time. At $t = 40 [s]$ two completely stable flows can be clearly observed, as shown in Fig. 2a in Guo et al. \cite{lftrans}.

\section{Conclusion}
We investigated the pattern formation in systems of active particles with chirality in the context of pedestrian flows. We considered the case when pedestrians move in a long corridor, such that half of them move in one direction and the other half move in the opposite direction. We performed extensive numerical simulations  of the system and provide a respective qualitative analysis. To describe the interactions between active particles (pedestrians), we  exploited the standard social force model and supplemented it with the simplest chirality force model. This force acts between approaching pedestrians moving towards each other and drives them to the right with respect to the direction of their motion. With increasing system  density, we observed the transition between disordered motion and multi-lane motion, as reported in previous studies, and quantified it in terms of the order parameter, introduced for systems of colloidal particles moving in an electrical field  \cite{LowenLanes2002}.  Additionally, we observed the transition from multi-lane to two-lane motion for non-vanishing chirality. This transition occurs either with increasing density for a fixed chirality or with increasing chirality for a fixed density. To describe these phase transitions, we developed mean-field qualitative theory. The theoretical estimates for the transition lines $\chi_*(\rho)$ and $\chi_{**}(\rho)$, on the phase diagram, which demarcate the disordered-multi-lane and multi-lane-two lane regimes, respectively, are in fair agreement with the simulation results. Finally, we demonstrated that the  presence of $10\%$ of pedestrians, with chirality opposite to that of the other $90\%$, does not destroy the two-lane state, although it initiates the emergence of additional narrow lines near the walls. In addition, we have qualitatively reproduced the experiments of Guo at all \cite{lftrans} in the ring-shaped corridor. Because left-right asymmetry is inherent in inter-pedestrian interactions, we believe that our model reflects an important feature of crowd dynamics. Hence, the present findings may be important for urban and transport modeling.

\section{Acknowledgement}

A.~S.~B. thanks Stefan~Kupper for fruitful discussions and Nikita~G.~Iroshnikov for assistance with the development of the simulation software.

%\bibliography{Pedestrians}

\end{document}